%% file: Moon_disk.tex
\documentclass[final,authoryear,3p,times,twocolumn]{elsarticle} 
\usepackage{amssymb}

\journal{Icarus}

\begin{document}

\begin{frontmatter}



\title{Investigation of the Initial State of the Moon-Forming Disk: Bridging SPH Simulations and Hydrostatic Models}


\author{Miki Nakajima}
\ead{mnakajima@caltech.edu}
\author{David J. Stevenson}
\address{Division of Geological and Planetary Sciences,
California Institute of Technology, 1200 E California Blvd., MC 150-21, Pasadena, CA 91125, USA.}

\begin{abstract}
\input{abstract}
\end{abstract}
\end{frontmatter}
%
\section{Introduction}
\label{intro}
\input{introduction}
%
\section{Model}
\label{model}
\input{model}
\section{Results}
\input{figures}
\label{results}
\input{results}
\section{Discussion}
\label{discussions}
\input{discussions}

\section{Conclusions}
\label{conclusions}
\input{conclusions}

\section*{Appendix}
\label{appendix}
\input{appendix}
\section*{Acknowledgement}
\label{acknowledgement}
\input{acknowledgement}

\bibliographystyle{elsarticle-harv}

\bibliography{miki.bib}







\end{document}

%% file: abstract.tex
According to the standard giant impact hypothesis, the Moon formed from a partially vaporized disk generated by a collision between the proto-Earth and a Mars-sized impactor. The initial structure of the disk significantly affects the Moon-forming process, including the Moon's mass, its accretion time scale, and its isotopic similarity to Earth. The dynamics of the impact event determines the initial structure of a nearly hydrostatic Moon-forming disk. 
However, the hydrostatic and hydrodynamic models have been studied separately and their connection has not previously been well quantified. 
Here, we show the extent to which the properties of the disk can be inferred from Smoothed Particle Hydrodynamic (SPH) simulations. By using entropy, angular momentum and mass distributions of the SPH outputs as approximately conserved quantities, we compute the two-dimensional disk structure. We investigate four different models: (a) standard, the canonical giant impact model, (b) fast-spinning Earth, a collision between a fast-spinning Earth and a small impactor, (c) sub-Earths, a collision between two objects with half Earth's mass, and (d) intermediate, a collision of two bodies whose mass ratio is 7:3. Our SPH calculations show that the initial disk has approximately uniform entropy. This is because the materials of different angular momenta are shocked to similar extents. The disks of the fast-spinning Earth and sub-Earths cases are hotter and more vaporized ($\sim$ 80-90\% vapor) than the standard case ($\sim$20\%). The intermediate case falls between these values. In the highly vaporized cases, our procedure fails to establish a unique surface density profile of the disk because the disk is unstable according to the Rayleigh criterion (the need for a monotonically increasing specific angular momentum with radius). In these cases, we estimate non-unique disk models by conserving global quantities (mass and total angular momentum). 
We also develop a semi-analytic model for the thermal structure of the disk, including the radial temperature structure and the vapor mass fraction. The model requires only two inputs: the average entropy and the surface density of the disk. 

%% file: introduction.tex
The widely accepted explanation for the origin of the Moon is the giant impact hypothesis, which involves a collision between the proto-Earth and an impactor during the late stage of terrestrial planet formation \citep{HartmannDavis1975, CameronWard1976}. 
A number of numerical simulations of giant impacts have been performed to test this hypothesis \citep{Benzetal1986, Benzetal1987, Benzetal1989, CameronBenz1991, CanupAsphaug2001, Canup2004, Canup2008}. 
Typically, a Mars-sized impactor hits the proto-Earth with a small impact velocity and large impact angle. 
The ranges of the impact velocity and angle are relatively limited because the angular momentum of the system is thought to be conserved over time \citep[e.g.,][]{Canupetal2001}. 
The typical outcome of these simulations is that the impactor is tidally disrupted, partly by tides and partly by the shock-induced flows,  and creates a massive iron-depleted circumplanetary disk, from which the Moon is subsequently accreted.
In this paper, we call this model the ``standard" scenario. This model can potentially explain several observed features, such as the Moon's mass and iron depletion, as well as the angular momentum of the Earth-Moon system. 

The standard model might fail to explain the observed isotopic similarities of the Earth and Moon. According to these simulations, most of the disk materials originate from the impactor, which is likely to have different isotope ratios from the proto-Earth. Then, the Moon should primarily inherit the isotopic signature of the projectile. This seemingly contradicts the nearly identical isotopic ratios of oxygen, silicon, tungsten, and titanium observed for the Earth and Moon \citep{Wiecheretal2001, Armytageetal2012, Toubouletal2007, Zhangetal2012}. A recent study shows that a giant impact with a smaller impact angle and larger impact velocity could lead to a disk that mainly originates from proto-Earth \citep{Reuferetal2012}, but this model still has difficulty with explaining such strong isotopic similarities.

\cite{PahlevanStevenson2007} suggest mixing have occurred between the disk and Earth's mantle through the connected atmosphere. 
This process could have homogenized the isotopic ratios, such as oxygen.
However, this model may have difficulty explaining the silicon isotopic ratios \citep{Pahlevanetal2011, Halliday2012}. 
It is also unclear whether the mixing is sufficiently efficient to accomplish the isotopic similarity even for oxygen, especially because this requires efficient homogenization all the way from the deep mantle of the Earth to the outermost half of the disk mass.

Recently, new dynamical models have been suggested for the origin of the Moon. \cite{CukStewart2012} propose a model in which an impactor hit a rapidly rotating, and hence oblate, proto-Earth (called the ``fast-spinning Earth"), whereas \cite{Canup2012} suggests a giant impact between two objects with half Earth's mass (hereafter ``sub-Earths"). In these models, the composition of the disk is similar to that of the Earth, so that the isotopic similarities can be potentially explained. In these studies, the angular momentum of the Earth-Moon system is approximately three times larger than its present-day value. \cite{CukStewart2012} suggest that the evection resonance between the Moon and the Sun can transfer excess angular momentum from the Earth-Moon system to the Sun-Earth. This is the resonance that occurs when the precession period of the Moon's pericenter is equal to the Earth's orbital period \citep{ToumaWisdom1998}. The efficiency of this angular momentum transfer may depend on a fortuitous choice of tidal parameters and their constancy with time, but we set aside this concern in this paper. 

To bridge such dynamical models and the resulting properties of the Moon, the thermodynamics of the Moon-forming disk needs to be understood.
The thermal structure of the disk affects the Moon-forming process and, hence, the chemical and isotopic compositions of the Moon.
As an example, the initial entropy of the disk may control the volatile content of the disk.
In a disk without radial mass transport or loss to infinity, the initial entropy will not matter since a high-entropy disk will simply cool to reach the same thermodynamic state as an initially low-entropy disk. However, the interplay of cooling and transport can be expected to affect the fate of the volatile components. This is important for explaining the volatile depletion of the Moon \citep{RingwoodKesson1977}.
Additionally, the isotope mixing occurs more efficiently in the high-entropy, and hence, vapor-rich disk \citep{PahlevanStevenson2007}.

The disk structure and its evolution have been studied analytically and numerically \citep[e.g.,][]{ThompsonStevenson1988, Idaetal1997, Kokuboetal2000, PritchardStevenson2000, GendaAbe2003, MachidaAbe2004, Ward2012, SalmonCanup2012}, but such previous studies are not directly connected with the giant impact modeling. Rather, these studies assume a circumplanetary disk as the initial condition. This discrepancy hinders deriving a realistic disk structure consistently. 

The aim of this paper is to derive, for the first time, the initial thermal structures of Moon-forming disks directly from giant impact simulations. First, we perform various giant impact simulations and then derive the hydrostatic disk structures based on the numerical results. Here we focus on four cases: (a) standard, (b) fast-spinning Earth, (c) sub-Earths, and (d) intermediate. Case (d) is  a collision of two bodies whose mass ratio is 7:3, which is similar to one of the calculations performed by \cite{Cameron2000}. Lastly, we explain a simple semi-analytic model that describes the thermal structure of the disk.

%% file: model.tex
The disk structure is derived in two steps.
First, giant impact simulations are performed using Smoothed Particle Hydrodynamics (SPH) \citep[e.g.,][]{Lucy1977, GingoldMonaghan1977, Monaghan1992}.
The details of SPH are summarized in Section \ref{SPH}. The endpoint of each simulation provides mass, angular momentum, and entropy distributions that form the starting point to determine the disk structure. 
\subsection{SPH integrated with GRAPE}
\label{SPH}
SPH is a Lagrangian method in which the fluid is modeled by numerous moving particles (grids). 
A particle $i$ has a mass $m_i$ and the so-called smoothing length, $h_i$.
The mass $m_i$ is distributed in a sphere of radius $2h_i$.
$h_i$ is defined such that approximately 50 neighboring particles are included in the sphere.
Increasing the number of particles near particle $i$ decreases $h_i$.
The momentum and energy equations are solved at each time step.
The momentum equation describes the forces due to gravity, pressure gradients, and shock compressions. 
The energy equation describes the change in internal energy due to shocks and adiabatic pressure work.  
The details have been described in previous studies \citep{MonaghanLattanzio1985, Monaghan1992, Canup2004}.
We have tested our SPH code by reproducing the analytical solution for a shock tube and the adiabatic collapse of an initially isothermal gas sphere \citep{Evrard1988, HernquistKatz1989}.
We use $N=100,000$ SPH particles, which is similar to modern SPH calculations \citep[e.g.,][]{CukStewart2012, Canup2012}. 
We have developed our own SPH code integrated with GRAvity PipE (GRAPE). The GRAPE hardware calculates gravitational interactions 100-1000 times faster than conventional computers at comparable cost \citep{Makinoetal1995, Makinoetal1997}. 
\subsection{Entropy of the disk}
\label{diskentropy}
In this work, entropy is used to characterize the thermodynamics of the system.  
Entropy is relatively well-conserved after the passage of shocks and is insensitive to the resolution.
This is because it slowly changes spatially, even when density and temperature change rapidly.
A particle with few nearby particles increases its smoothing length and experiences artificial adiabatic expansion and cooling (without its neighboring particles, the density of a particle $i$ is $\rho_i \propto \ m_i/h_i^3 $). This effect can lead to unrealistically a small density and temperature of the particle.
Because a particle in the disk does not have sufficient neighboring particles with the present-day resolution ($N \sim10^5$), it experiences significant adiabatic expansion. This leads to the unrealistically small density and temperature, while the entropy does not change in the process. 

The majority of the entropy gain is due to the impact-induced shocks, which can be modeled by SPH simulations.
Additionally, we allow for the small increments in entropy arising from SPH particles that are initially in eccentric or inclined orbits but are incorporated into the disk as material in circular orbit. Entropy gain due to mass redistribution within the disk is also possible, even on a short timescale. This effect is considered for high entropy (high vapor fraction) cases. The detailed ideas and procedures are described in Section \ref{model:disk} and the Appendix.

\subsection{Initial conditions}
\label{initial}
Both the target and impactor consist of 70\% mantle (forsterite) and 30\% core (iron) by mass. 
The M-ANEOS equation of state is used \citep{ThompsonLauson1972, Melosh2007}. 
This equation can treat phase changes and co-existing multiple phases.
Initially, the mantle has a uniform entropy of 3165 J/K/kg, such that temperature at the surface is approximately 2000K. 
The surface temperature corresponds to that of the ``warm starts" of the previous studies \citep{Canup2004, Canup2008}.
The constant entropy at depth implies a solid that is close to the melting curve. This is appropriate since the early Earth was formed so quickly that it could not have efficiently cooled in the period since the previous giant impact. 
The parameters are   
the impactor-to-total mass ratio $\gamma$,  
the total mass of the target and impactor, $M_{\rm T}$, 
the scaled impact parameter $b$ ($b=\sin \theta$, here $\theta$ is the impact angle),
the impact velocity $v_{\rm imp}$,
and the initial spin period, $T_{\rm spin}$.  

\subsection{Derivation of the disk structure}
\label{model:disk}
The mass and entropy of the disk is obtained from an SPH simulation.
After approximately one day of simulated time, the system has usually evolved to the point where its final state can be estimated from the current distribution of orbital parameters and thermodynamic state. 
The semi-major axis, eccentricity, and inclination of each particle orbiting around the planet are identified based on the SPH output.
A particle whose orbit is not hyperbolic or does not cross the planet is considered to be a part of the disk. The details are described in a previous study \citep{Canupetal2001}.
Additionally, we also detect aggregates of SPH particles and assign a single orbit to them.
Assuming the inclination and eccentricity are quickly damped \citep{ThompsonStevenson1988} while the angular momentum is conserved, the corrected semi-major axis of a particle $i$, $a_{i, \rm final}$, is obtained. The details are described in the Appendix. 
The entropy of the particle, $S_i$, is directly obtained from the SPH calculation.
Based on $m_i$ and $S_i$ as a function of $a_{i, \rm final}$, 
the cylindrically averaged surface density of the disk, $\Sigma(r)=dM/2\pi r dr$, and entropy, $S_{\rm SPH}(r)$, are derived. Here, $r$ is the distance from the Earth's spin axis and $dM$ is the mass between the two cylindrical shells.

A problem occurs with this approach when the vapor mass fraction is high. Clearly, SPH particles that will expand mostly to vapor will not follow approximately Keplerian trajectories due to the large pressure gradient forces and possibly instabilities as explained below. 
The vapor-rich disk is partially supported by the pressure gradient ($dp/dr$).
When the self gravity of the disk is negligible, the specific angular momentum of the vapor in the $z$-direction
(parallel to the Earth's spin axis), $L_z$, is written as
\begin{equation}
L_z(r)=\sqrt{GM_{\rm p}r+\frac{r^3}{\rho} \frac{dp}{dr}},
\label{eq:kepler}
\end{equation}  
where $M_{\rm p}$ is the mass of the planet.
In the outer part of the disk, $dL_z/dr$ becomes negative since the term that includes $r^3 dp/\rho dr$ ($<0$) starts dominating the $GM_{\rm p}r$ term.
These regions do not satisfy the Rayleigh stability criterion ($dL_z/dr>0$, \citealp{Chandrasekhar1961}) and will mix radially on a dynamical timescale (i.e., of order hours). 
In addition, if $d\Sigma/dr>0$ at the inner edge of the vapor-rich disk, there will be an radial mass redistribution, and some vapor will migrate inward and merge with the planet due to the pressure gradient forces.

Therefore, it is not possible to determine the surface density in these cases using only the SPH output and the procedure for treating the SPH particles as Keplerian. In principle, a sufficiently high-resolution SPH simulation carried out to longer times could determine the structure. We have chosen instead to estimate the outcome by use of a simple functional form that satisfies the global constraints. 
We assume that the surface density can be written as $\Sigma(r)=(C_1 + C_2r)\exp(-C_3r)$ ($C_1, C_2$, and $C_3$ are constants). These constants are obtained by conserving $M_{\rm D}$ and $L_{D}$ and meeting a stable condition that $d\Sigma/dr=0$ at the inner edge (This choice of boundary condition is not physically required and other choices could be possible). 
The resulting disk is more stable and meets the Rayleigh criterion in the broader regions. However, the disk may be still unstable near the outer edge because $L_z(r)$ rapidly decreases at large $r$, but this does not cause a major problem because the mass present in the outer region is relatively small.

Next, the two-dimensional structure of the disk is calculated based on the following assumptions: 
(1) The disk is hydrostatic in the $z$-direction.
(2) A vapor phase exists above a thin liquid layer (if it exists) and no mixed-phase layer exists. 
(3) The gravity forces of the planet and liquid of the disk are considered. 
(4) If a liquid layer exists at $z=0$ at a given $r$, the pressure is equal to the saturation vapor pressure at any $z>0$. 
Otherwise, the pressure gradient follows a dry adiabatic lapse rate until the pressure reaches the saturation vapor pressure. The vapor phase above is also saturated.   
Whether a liquid layer exists at a given $r$ depends on $\Sigma(r)$ and $S_{\rm SPH}(r)$.  The details are described in Section \ref{simplemodel}. 
The hydrostatic relation can be described as 
\begin{equation}
\frac{1}{\rho}\frac{dp}{dz}=-\frac{GM_{\rm p}z}{(r^2+z^2)^{\frac{3}{2}}}-2\pi G\Sigma_{\rm l}(r), 
\label{eq:hydro}
\end{equation} 
where $p$ is the pressure, $\rho$ is the density, $G$ is the gravitational constant, $M_{\rm p}$ is the mass of the planet, and $\Sigma_{\rm l}(r)$ is the surface density of the liquid at a given $r$.

The thermodynamic properties, such as $p$, $\rho$, and the temperature, $T$, are iteratively calculated in the $z$-direction by solving Equation (\ref{eq:hydro}), while satisfying $\Sigma(r)$ and $S_{\rm SPH}(r)$. 
Based on an initial guess of $T(r,z=0)$ at a given $r$, the entropy averaged in the $z$-direction $S(r)$ is calculated using the relationship 
\begin{equation}
\Sigma (r) S(r)=\int_{-\infty}^{\infty}\rho_{\rm v}(r,z)S_{\rm v}(r,z)dz+\Sigma_{\rm l}(r)S_{\rm l}(r).
\label{eq:entro_mass}
\end{equation}  
Here, $\rho_{\rm v}(r,z)$ and $S_{\rm v}(r,z)$ are the density and entropy of the vapor, $\Sigma_{\rm l}(r)(=\Sigma(r)-\Sigma_{\rm v}(r))$ and $S_{\rm l}(r)$ are the surface density and entropy of the liquid. By comparing $S_{\rm SPH}(r)$ with $S(r)$, the temperature at the interface ($z=0$) is corrected until the relative change of these two values becomes less than 0.1\%.

%% file: figures.tex
\begin{center}
\begin{table*}[ht]
{\small
\hfill{}
\begin{tabular}{|l|c c c c c|| c c  c c |}
\hline
\textbf{}& $\gamma$ & $M_{\rm T}$ & $b$ & $v_{\rm imp}$&$T_{\rm spin}$ & $M_{\rm D}$ & $L_{\rm D}$ & $S_{\rm ave}$ & VMF  \\
\hline
(a) Standard &0.13&1.02&0.75&1.0&0&1.35&0.26&4672&19\%\\
(b) Fast-spinning Earth &0.045&1.05&-0.3&20 (km/s)&2.3&2.36&0.44&7132* &96\%\\  
(c) Sub-Earths &0.45&1.04&0.55&1.17&0&3.07&0.64&7040*&88\%\\ 
(d) Intermediate &0.3&1.00&0.6&1.0&0&2.80&0.57&5136&31\%\\
\hline
\end{tabular}}
\hfill{}
\caption{The initial conditions and outcomes. $\gamma$ is the impactor-to-total mass ratio, $M_{\rm T}$ is the total mass scaled by the Earth mass, $b$ is the scaled impact parameter, $v_{\rm imp}$ is the impact velocity scaled by the escape velocity (except b), and $T_{\rm spin}$ is the initial spin period of the target (hrs). $M_{\rm D}$ is the disk mass scaled by the Moon mass, and $L_{\rm D}$ is the angular momentum of the disk scaled by the current angular momentum of the Earth and Moon. $S_{\rm ave}$ is the averaged entropy of the disk (J/K/kg). The asterisk indicates that the entropy increase due to mass-redistribution of the disk is considered (discussed in Section \ref{model:disk} and the Appendix.)
VMF is the vapor mass fraction. }
\label{tb:SPHsummary}
\end{table*}
\end{center}
\begin{figure*}
  \begin{center}
     \includegraphics[scale=1.0]{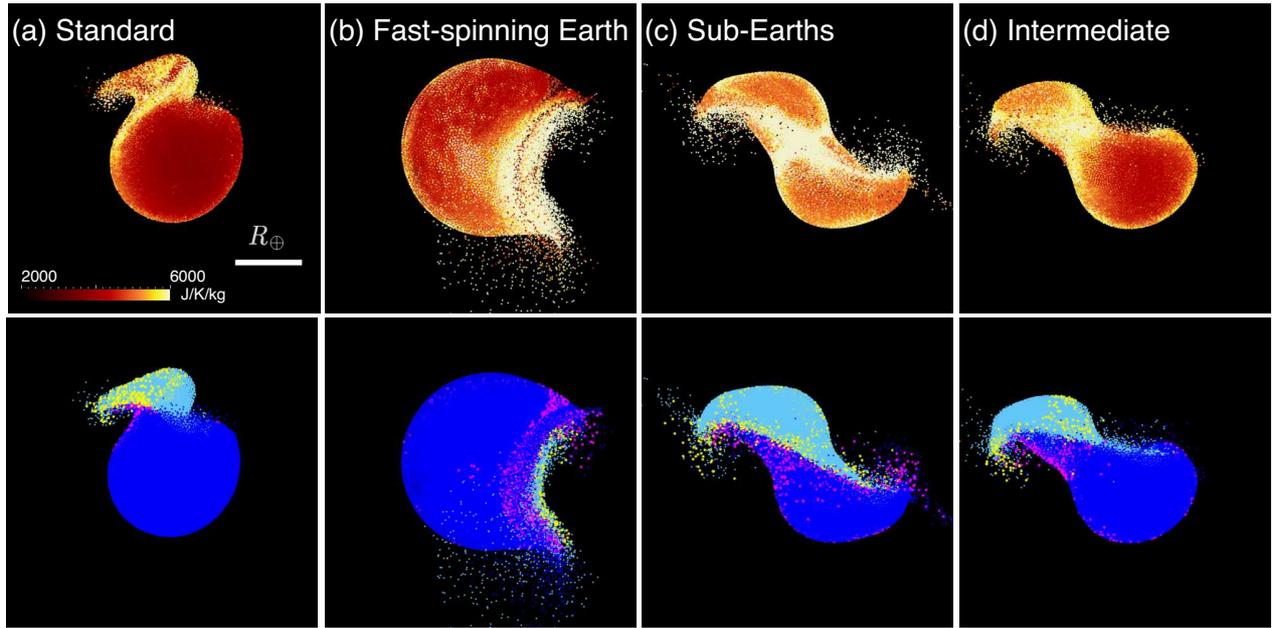}
  \end{center}
  \caption{Each panel shows a projection of a 3D calculation onto the equatorial plane.
In the top panel, color scales with the entropy of forsterite in J/K/kg.
In the bottom panel, particles originating from the target are shown in blue and magenta.
Particles originally from the impactor are shown in sky blue and yellow. 
The magenta and yellow particles (their sizes are magnified) become part of the disk. 
 }
\label{fig:origins}
\end{figure*}
\begin{figure*}
  \begin{center}
    \includegraphics[scale=1.0]{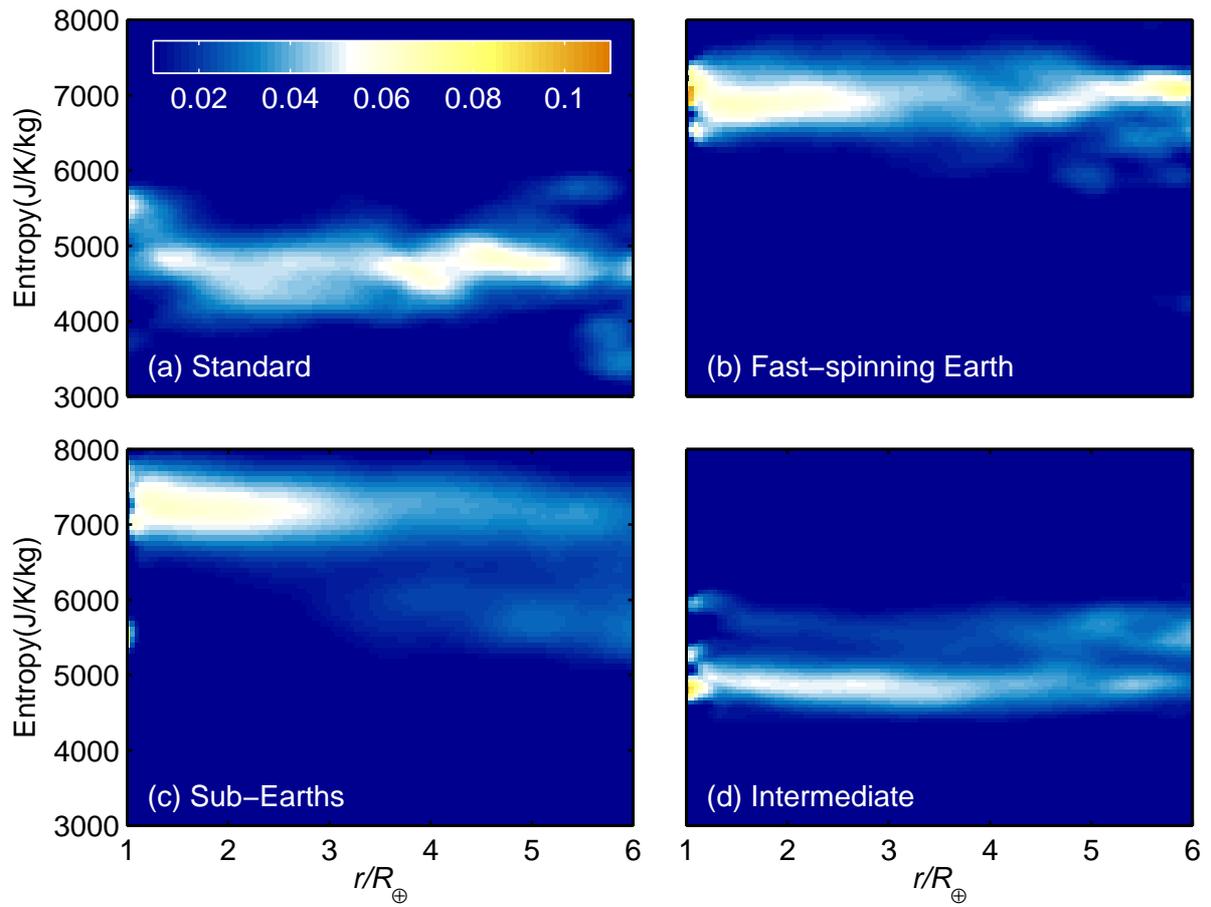}
  \end{center}
  \caption{
A probability distribution of a particle, P$(r,S)$, where $r$ is a distance from the Earth's spin axis and $S$ is entropy.
The probability is normalized at a given $r$ ($\Sigma_k {\rm P}(r,S_k)=1$) and color-coded according to its intensity. 
P$(r,S)$ is a weak function of $r$ and the disk is nearly isentropic in all cases. 
}
\label{fig:entro}
\end{figure*}
\begin{figure*}
  \begin{center}
    \includegraphics[scale=1.0]{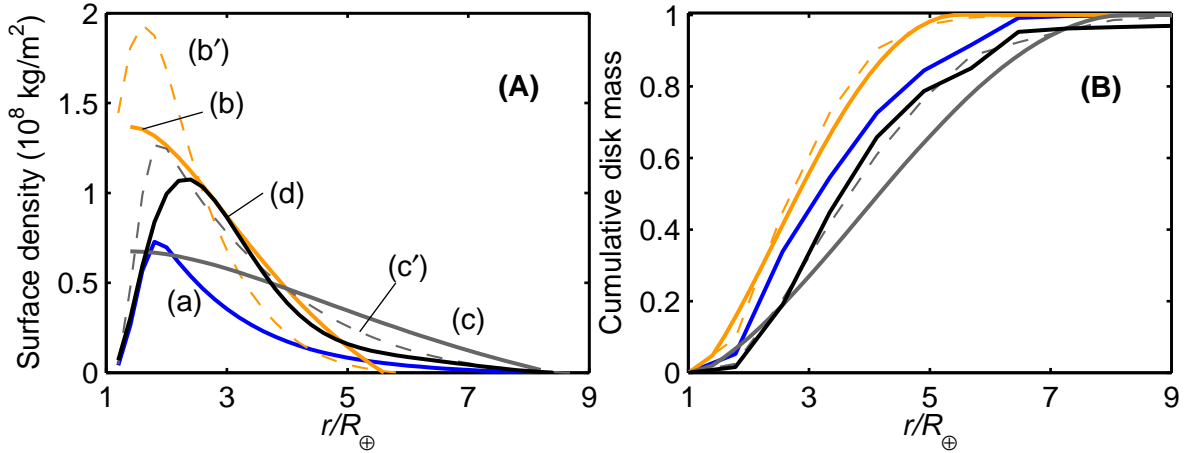}
  \end{center}
  \caption{The surface density and cumulative mass distribution of the disk. 
(a) standard: blue, (b) fast-spinning Earth: yellow, (c) sub-Earths: gray, (d) intermediate: black.
(A) The SPH calculations produce the surface densities of (a), (b'), (c'), and (d).
However, (b') and (c'), indicated by the dashed lines, are unstable because these disks do not satisfy the Rayleigh criterion. 
The surface densities of more stable disks are shown in (b) and (c).
(B) The mass distribution is obtained by integrating the surface density. 
  }
\label{fig:mass_surf}
\end{figure*}
\begin{figure*}
  \begin{center}
   \includegraphics[scale=1.0]{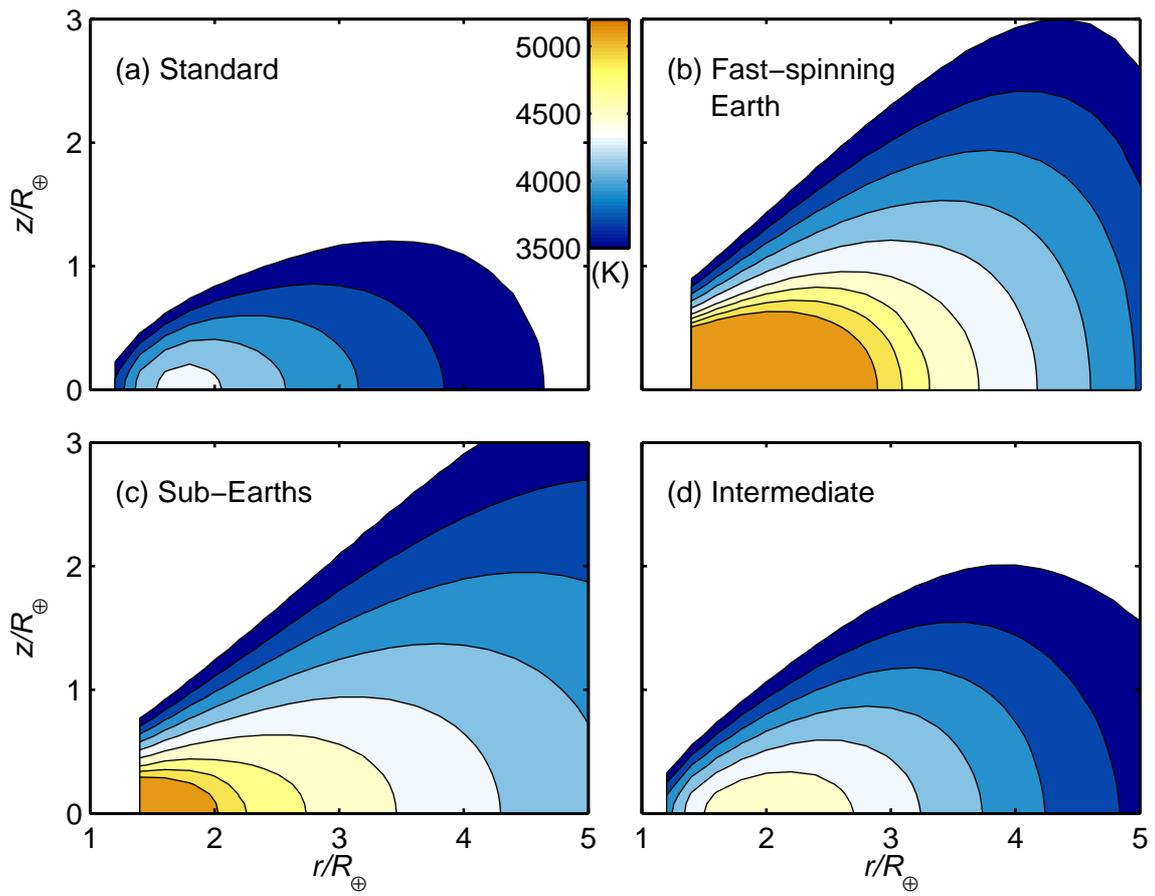}
  \end{center}
  \caption{Temperature contours of the disk. More energetic impacts lead to higher disk temperatures. 
  In (b) and (c), the inner edge of the disk is located slightly outside of the disk in (a) and (d) because its planetary equatorial radius is larger due to its short spin period. }
\label{fig:2D}
\end{figure*}
\begin{figure*}
  \begin{center}
   \includegraphics[scale=1.0]{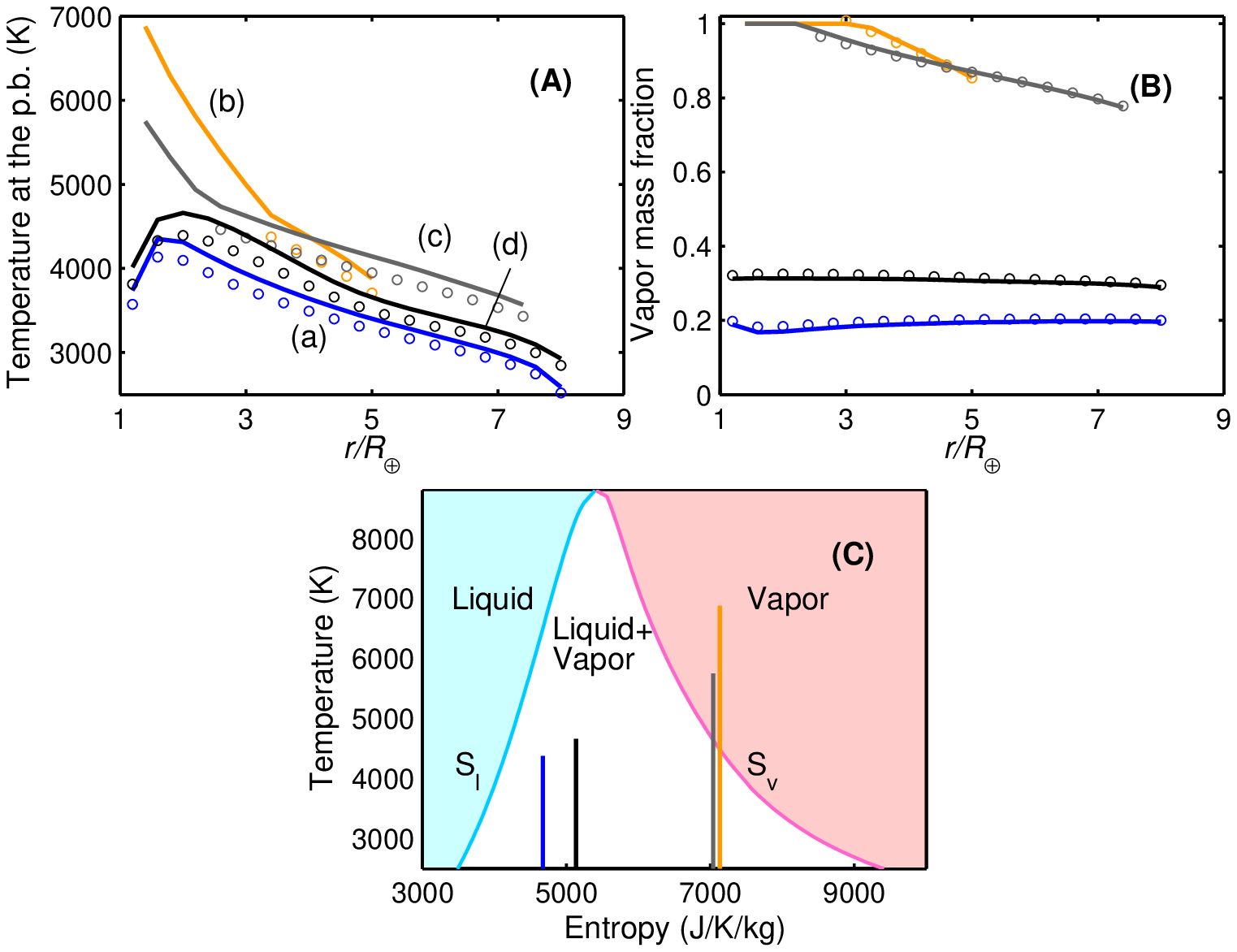}
  \end{center}
  \caption{
 (A) and (B) show the comparisons between the numerical result (line) and the simple model (circle) described in Section \ref{simplemodel}. 
 The color-coding of the lines is the same as that of Figure \ref{fig:mass_surf}.
 (A) shows the temperature at the phase boundary (at $z=0$). 
 The difference of $\sim$200 K between the simulation and model arises because the self-gravity of the disk and vertical variation of the temperature are ignored in the model.
The peaks of the surface density and temperature coincide. 
 (B) shows the radial vapor mass fraction. The model matches the numerical result well.  
 (C) is the phase diagram of forsterite (M-ANEOS, \citealp{ThompsonLauson1972, Melosh2007}).
The sky-blue solid line represents the entropy of liquid and the magenta solid line represents that of the vapor at the phase boundary.
The range of $T_{\rm pb}$ for each $S_{\rm ave}$ is indicated by the solid lines. 
}
\label{fig:temp_vapor}
\end{figure*}

%% file: results.tex
The initial conditions and outcomes are summarized in Table \ref{tb:SPHsummary}. $M_{\rm D}$ is the disk mass normalized by the lunar mass and $L_{\rm D}$ is the disk angular momentum normalized by that of the Earth-Moon system. $S_{\rm ave}$ is the average disk entropy (discussed in Section \ref{isen}) and determined by the impact energy and material properties.
 VMF is the overall vapor mass fraction of the disk.
The $S_{\rm ave}$ of the fast-spinning Earth (b) and sub-Earths (c)  is much larger than that of standard (a) due to the larger kinetic energy involved in the impact.
This high-entropy disk is mainly vapor (discussed in Section \ref{simplemodel}). The disk in the intermediate (d) is moderately shock-heated.
The entropy increase due to the circularization of the particles in the disk is typically less than 10\%.
The mass redistribution is considered in (b) and (c), which also increases $S_{\rm ave}$ (indicated by the asterisk). The value of this increment depends on the model of the surface density, but it is not significant (less than 5\%). The details are discussed in Section \ref{model:disk} and the Appendix.
The disk of the non-standard models is more massive than that of the standard model.
After the circulation of the disk, the mass outside of the Roche radius ($\sim 2.9R_{\oplus}$) is (a) 0.77, (b) 1.12, (c) 2.33, and (d) 1.96 in lunar mass. 

\subsection{Isentropic disk}
\label{isen}
Figure \ref{fig:origins} shows snapshots of the impact simulations of the four different scenarios. 
The upper panels show the entropy at the impact and the bottom panels show the origin and fate of the particle.
The magenta particles (target-origin) and yellow particles (impactor-origin) become part of the disk. 
We refer to the particles that eventually form the disk as ``disk particles" hereafter. 
In (a), the impactor is destroyed by the impact and the tides from the planet and then form a disk.
Most of the disk particles come from specific parts of the impactor. 
Since the disk particles have similar distances from the impact point,  they gain similar extent of shock-heating and entropy. 
This leads to a relatively isentropic disk.
This feature is clearly shown in Figure \ref{fig:entro}a. 
This figure shows the probability distribution of a disk particle, ${\rm P} (r,S)$, color-coded according to its intensity.
The number of the SPH particles is counted at a given $r \pm \Delta r$ and $S \pm \Delta S$ ($\Delta r=0.02 R_{\oplus}$ and $\Delta S=20$ J/K/kg). 
${\rm P} (r,S)$ is obtained by normalizing the number by the total number of SPH particles at that radius.
Note that ${\rm P} (r,S)$ is statistically irrelevant in a sparse region (near the outer edge of the disk).
Although the entropy of the disk in (a) has a dispersion, it is relatively uniform and a weak function of $r$.

In (b), a small impactor smashes into the rapidly-rotating oblate planet.
The impact is so energetic that a part of the planetary mantle is stripped off. 
Most of the disk particles are initially located near the surface (shown in magenta in Figure \ref{fig:origins}b).
These particles are ejected in the $z$-direction after the first shock.
When they fall back to the $z=0$ plane, they collide with the particles coming from the other side.
This secondary shock significantly increases the entropy of the disk particles, because the pre-shock density of the second impact is small and the disk materials are compressed easily.
Since most of the disk particles follow this evolutionary path, the disk particles are shock-heated to a similar extent (Figure \ref{fig:entro}b).
In (c), the two similar-size objects collide with each other several times until they merge into a single planet. 
The angular momentum of the planet becomes so large that the planet becomes unstable. 
A portion of the outer mantle is stripped away and forms a disk. 
The entropy of the disk particles continue to increase through the multiple impacts.
This leads to a larger dispersion of disk entropy than the other cases (Figure \ref{fig:entro}c), but the entropy still does not depend on $r$. 
(d) is similar to (a), but disk materials are coming from broader regions, including the target.
Nevertheless, the disk particles are originally located nearby and experience similar increase of the entropy (Figure \ref{fig:entro}d).

Thus, the entropy is a weak function of $r$ and approximately uniform in all cases.
In the following arguments we assume the disk is adiabatic and has a uniform entropy ($S_{\rm SPH}(r)=S_{\rm ave}=$ const.). 
Note that $S_{\rm SPH}$ includes additional heating due to the circularization of the disk particles (Section \ref{model:disk} and the Appendix), although the increments are relatively small (typically, less than 10\% of $S_{\rm ave}$). 
For (b) and (c), additional heating is considered due to the mass redistribution in the disk (Section \ref{model:disk} and the Appendix). 
\subsection{Radial mass distribution}
\label{radialmass}
Figure \ref{fig:mass_surf}A shows the surface density of the disk. 
The blue solid line (a), the yellow dashed line (b'), the gray dashed line (c'), and the gray solid line correspond to the surface densities of the cases (a), (b), (c), and (d) based on the SPH calculations. 
The disk in (b') and (c') is unstable because the Rayleigh criterion is not satisfied at $r>3.4 R_{\oplus}$ and $ r>4.8 R_{\oplus}$ for (b') and (c'), respectively.
Even in the stable regions, the slope of $L_{\rm z}$ is shallow, especially for (b'), so that the regions may become easily unstable. 
Additionally, since $d\Sigma/dr>0$ at the inner edge of the disk, the mass is redistributed radially.

The surface density of a more stable disk ($\Sigma(r)=(C_1 + C_2r)\exp(-C_3r)$) is shown as (b) and (c).
This modeled disk is stable in a broader region, but still unstable at $r>3.8R_{\oplus}$ and $r>5.4 R_{\oplus}$ in (b) and (c). 
However, the slope of $L_z$ of (b) and (c) is steeper than that of (b') and (c'), so that the disk is more stable. 
The disk in (a) and (d) does not satisfy the criteria either and hence its structure may be reshaped. 
But we ignore this change because the disk in (a) and (d) is liquid-dominant and the effect of pressure gradient is less significant.

Figure \ref{fig:mass_surf}B shows the cumulative disk mass. This is obtained by integrating the surface density (Figure \ref{fig:mass_surf}A).
The disk in (b) is more compact than in the other cases, because the disk materials originate from the mantle of the target and they gain relatively little angular momentum from the impact. In the other cases, the angular momentum comes from the orbital angular momentum of the massive impactor. 
%
%
%
\subsection{2D structure of the disk}
The temperature is derived at a given $r$ and $z$ from the entropy and phase diagram. 
It is important to stress that entropy is a better parameter to characterize the thermodynamics of the disk, but temperature is of course is important to decide chemical questions such as partitioning and possible isotopic fractionation. 
The temperature is also relevant to the evaluation of possible hydrodynamic outflow.
Figure \ref{fig:2D} shows the temperature contour of the disks as a function of $r$ and $z$, normalized by $R_{\oplus}$.
The figure is color-coded according to the temperature (3500-5200K). 
A more energetic impact (with higher values of $S_{\rm ave}$) leads to higher temperatures in the disk.
\subsection{Simple semi-analytic disk model}
\label{simplemodel}
The structure of the disk can be described by a simple model. 
Equation (\ref{eq:entro_mass}) can be approximately written as
\begin{equation}
\Sigma (r) S_{\rm ave}  \simeq \Sigma_{\rm v}(r) S_{\rm v}(r,0)+\Sigma_{\rm l}(r)S_{\rm l}(r),
\label{eq:approx1}
\end{equation}  
where, $S_{\rm SPH}(r) = S(r) = S_{\rm ave}$ is assumed. $S_{\rm v}(r,0)$ is the entropy at the phase boundary ($z=0$) at a given $r$.  This approximation is reasonable because the entropy variation in the $z$-direction is relatively small. Additionally, because the density is largest at $z=0$, $S_{\rm v}(r,0)$ has the largest contribution to $S (r)$ among $S_{\rm v}(r,z)$. Neglecting the self gravity of the disk and assuming that the temperature variation in the $z$-direction is small, $\Sigma_{\rm v}$ is written as
\begin{equation}
\Sigma_{\rm v}(r) = \int_{-\infty}^{\infty}\rho_{\rm v}(r,z)dz \sim \int_{-\infty}^{\infty}\rho_{\rm v}(r,0) e^{-(z/H)^2}dz,
\label{eq:approx2}
\end{equation}
which becomes
\begin{equation}
\Sigma_{\rm v}(r)  \sim \sqrt{\pi} \rho_{\rm v}(r,0)H,
\label{eq:approx2}
\end{equation}
as previously derived \citep{ThompsonStevenson1988, Ward2012}.
Here, $H=\sqrt{2}c/\Omega$, $c=\sqrt{RT(r,0)/\mu}$, and $\Omega=\sqrt{GM_\oplus/r^3}$. $\mu$ is the average molecular weight ($\sim 30$ g/mol, \citealp{ThompsonStevenson1988}). Likewise, the pressure distribution is also derived as $p(r,z)\sim p(r,0)\exp{[-(z/H)^2]}$.

Equation (\ref{eq:approx1}) can be written as
\begin{equation}
S_{\rm ave}  \sim f S_{\rm v}(r,0)+(1-f)S_{\rm l}(r),
\label{eq:approx3}
\end{equation}  
\begin{equation}
f(r) \equiv \frac{\Sigma_{\rm v}(r)}{\Sigma(r)},
\label{eq:approx4}
\end{equation}  
where $f(r)$ is the vapor mass fraction at a given radius $r$. 

Note that all parameters on the right hand side of Equation (\ref{eq:approx3}) depend only on the temperature at  the phase boundary, $T_{\rm pb}(r) (=T(r,0))$. 
Since the liquid and vapor are assumed to be in equilibrium, $\rho_{\rm v}(r,0)$ and hence $\Sigma_v(r)$ are uniquely determined, once $T_{\rm pb}(r)$ is specified. 
$T_{\rm pb}$ is approximately obtained from Equation ($\ref{eq:approx3}$), given $\Sigma(r)$, $S_{\rm ave}$, and an equation of state.
Replacing $r$ by $T_{\rm pb}$ ($r=r(T_{\rm pb})$), $f$ is written as
\begin{equation}
f(T_{\rm pb}(r)) \sim \frac{S_{\rm ave}-S_{\rm l}(T_{\rm pb})}{S_{\rm v}(T_{\rm pb})-S_{\rm l} (T_{\rm pb})}.
\label{eq:vapor_model}
\end{equation}   

Figure \ref{fig:temp_vapor} shows the validity of this model. In Figure \ref{fig:temp_vapor}A, the $T_{\rm pb}$ calculated numerically by solving Equation (\ref{eq:hydro}) are indicated by the lines and the model (the approximate $T_{\rm pb}$ by solving Equation ($\ref{eq:approx3}$)) is indicated by the circles. 
A systematic difference ($\sim 200$K) occurs because the temperature variation in the $z$-direction and the self-gravity of the disk are ignored here, but the model still captures the behavior. 
Figure \ref{fig:temp_vapor}B shows radial vapor mass fraction. The model matches the calculation very well. 
While the vapor fraction does not change radially in (a) and (d), it does change in (b) and (c). 
Additionally, the inner part of the disk in (b) and (c) is completely in the vapor phase.
This vapor-only region cannot be modeled since the pressure is not equal to the saturation vapor pressure.
Figure \ref{fig:temp_vapor}C provides visual interpretations of Equation (\ref{eq:vapor_model}).
The shaded regions in sky-blue and magenta show the liquid and vapor phases of forsterite.
$S_{\rm ave}$ of each disk is represented by the solid line. 
$f=(S_{\rm ave}-S_{\rm l})/(S_{\rm v}-S_{\rm l})$ does not vary greatly in (a) and (d) as in the other cases.
Our model also provides an intuition for the reason that a higher $\Sigma(r)$ tends to give a higher $T_{\rm pb}$ (Figure \ref{fig:mass_surf}A and \ref{fig:temp_vapor}A).
$\Sigma_{\rm v}(r)$ is basically determined by $T_{\rm pb}$. For a given $r$ and $T_{\rm pb}$, a higher $\Sigma(r)$ results in a smaller $S(r)$, since the contribution of $\Sigma_{\rm l} S_{\rm l}$ becomes larger in Equation (\ref{eq:approx1}) ($S_{\rm v} \geq S_{\rm l}$).
Therefore, given a higher $\Sigma(r)$, $T_{\rm pb}$ increases to obtain the same $S(r)$.  
Additionally, given $\Sigma(r)$, a smaller $r$ leads to a smaller $\Sigma_{\rm v}(\propto H \propto r^{3/2})$. Thus, to obtain the same $S_{\rm ave}$, $T_{\rm pb}$ needs to be greater at a smaller $r$.

The vapor mass fraction in (b) and (c) reaches 100\% at the inside of the disk. 
In this region, both $\Sigma(r)$ and $S_{\rm ave}$ are large.
To reach such high $S(r)(=S_{\rm ave})$, $T_{\rm pb} (r)$ needs to be large. 
However, an upper limit of $T_{\rm pb} (r)$, and hence $S(r)$, exists to satisfy $\Sigma_{\rm v} (r) \leq \Sigma(r)$. 
Let these limits called be $T_{\rm pb(max)}(r)$ and $S_{\rm max}(r)$.
$dT_{\rm pb(max)}(r)/d\Sigma(r)>0$ because a larger $\Sigma(r)$ allows a larger $T_{\rm pb}$.
Near this limit, $f$ is close to unity and $S(r)$ should be close to $S_{\rm v}(r)$. 
Since $dS_{\rm v}(T)/dT <0$ according to the phase curves of forsterite (Figure \ref{fig:temp_vapor}C), $dS_{\rm v}(r)/d\Sigma(r) \sim dS_{\rm max}(r)/d\Sigma(r)<0$. 
Therefore, a given $S_{\rm ave}$ may not be satisfied if $\Sigma(r)$ is too large in the region. 
In such a region, the assumption that the system is in vapor-liquid equilibrium is no longer valid.
The part of the disk is completely in the vapor phase.
This is the case for the inner region of (b) and (c). 
Once $T_{\rm pb}(r)$ is determined, $\rho(r,z)$ and $p(r,z)$ are approximately obtained, as discussed ($\rho(r,z)\sim \rho(r,0)\exp{[-(z/H)^2]}$, $p(r,z)\sim p(r,0)\exp{[-(z/H)^2]}$).
Although this model is very simple, it is useful in that the thermal structure of the disk can be semi-analytically approximated based on just two parameters $S_{\rm ave}$ and $\Sigma(r)$.

%% file: discussions.tex
\subsection{Structure of the disk}
In (b) and (c), the modeled surface density is not uniquely determined by the SPH output (Section \ref{model:disk}), but this is only a problem in the high vapor fraction cases.
A different model of the surface density will provide different $S_{\rm ave}$ according to the potential energy differences, $\Delta U$, of the disks (Appendix).
The increment of $S_{\rm ave}$ due to $\Delta U$ is typically a few percent. Therefore, although the disk model is not unique, our model still describes the general disk structure.

Here, we only consider one giant impact simulation for each model. 
However, the surface density and entropy of the disk can vary even in the same type of the giant impact model.
Previous statistical studies show that disk mass and angular momentum vary, depending on the choices of initial conditions.
This is the case especially for grazing impacts \citep[e.g.,][]{Canup2004}.
Therefore, the initial conditions affect the structure of the disk (personal communications with Kaveh Pahlevan).

Note that the atmosphere of the planet has not been considered. 
Because the SPH method does not adequately describe the physics at the interfaces of large density differences, it cannot very well describe the planet-atmosphere or core-mantle (iron-forsterite) boundaries. 
This limitation arises because the density of an SPH particle $i$ is determined by its neighboring particles.
If the neighboring particles have much larger or smaller densities than that of the expected density of the particle $i$, 
it leads to an artificially large or small density of the particle.
At the end of an SPH simulation, the outer part of the planet is inflated and has an atmosphere-like structure. 
But the mass and size of the atmosphere may include large errors.
For simplicity we exclude the effect of this ``atmosphere" and merely consider it part of the planet.

\subsection{Stability of a vapor-rich disk}
It has been suggested that a vapor-rich disk is not suitable for the Moon formation.
\cite{Wadaetal2006} perform two grid-based hydrodynamic simulations of a standard-type giant impact.
They use two different polytrope-type equations of state (EOS) and compare the outcomes. One of the EOSs mimics a ``gas-like" material and the other represents a liquid (or solid) material.
They find out a giant impact with the gas-like EOS leads to a dynamically unstable disk.
The density contrasts within the disk are so large that several shocks propagate through the disk. 
The disk loses its angular momentum by the process and a significant disk mass falls onto the planet. 
On the other hand, if the disk is mostly liquid (i.e., with the liquid-like EOS),
 the density contrast within the disk is small, so that such strong shocks are not created. 
They conclude that a vapor-rich disk loses its significant mass so quickly that it is not suitable for the Moon formation.
This is an interesting outcome, but it is unclear in their result exactly how the angular momentum budget is satisfied (since material falling back onto Earth is necessarily balanced by material that gains angular momentum). 
Additionally, it is uncertain whether this is applicable to the stability of the vaporous disk in (b) and (c).
The polytrope-type EOSs cannot describe the behavior of the realistic mantle materials very well, such as the phase changes.
Therefore, the density structure of the disk may not be very physical. 
In addition, the proposed EOS may not provide realistic estimates of the Mach numbers of the shocks either.
This may provide a large uncertainty in estimating the loss of the angular momentum and mass of the disk due to the shock passage.
In order to understand the stability of the disk, a more realistic EOS needs to be implemented with such a grid-based code.

\subsection{Evolution of the disk}
The Moon-forming disks of the fast-spinning Earth (b) and sub-Earths (c) have much higher entropy and vapor mass fractions than those of the standard model (a).
The evolution of such a highly shock-heated disk would likely follow a different path from the canonical scenario.
The extent of vaporization may affect the chemical and isotopic signature of the disk and the resulting Moon \citep{Pahlevan2013}. 
Additionally, the mass of the Moon might become an important constraint.
It takes a longer time to radiatively cool this high-entropy disk. 
If the disk experiences viscous spreading during the cooling process, it may lose its mass to the planet even before condensation to allow Moon formation.
In addition, because the gas is orbiting more slowly than the liquid, the gas removes the angular momentum of the liquid. 
Thus, the liquid droplets fall to the planet while the gas moves outward, leading to the additional mass loss of the disk.
However, the effect may not prevent a Moon formation, because the disk of (b) and (c) is more massive than (a).
In any event, further study is required for a more quantitative argument. 

Our study also shows that the vapor fraction of the disk can radially vary in (b) and (c), which may cause some chemical and isotopic heterogeneity in the disk. 
If the Moon (or at least its surface) preferentially formed from a particular location of the disk \citep{SalmonCanup2012}, the Moon's chemical and isotopic signatures might not represent the entire disk, but rather a specific region of the disk. 
However, this conclusion depends on the efficiency of the radial mixing.

\subsection{Effects of the equation of state}
The forsterite M-ANEOS has several non-negligible caveats. 
Recent experimental studies indicate that M-ANEOS underestimates the shock-induced entropy gain of silicates \citep{Kurosawaetal2012, Krausetal2012}. 
\cite{Kurosawaetal2012} suggest that this effect becomes prominent when the peak pressure, $P_{\rm peak}$, is higher than 330 GPa. 
Although this may lead to the higher entropy and vapor-mass fraction of the disk, the increments may be relatively limited. 
This is because majority of the disk particles have smaller peak pressures than this.
The fraction of disk particles with $P_{\rm peak} > 330$ GPa is (a) 0 \%, (b) 30 \%, (c) 6.5 \%, and (d) 0 \%. 
However, the entropy gain by an impact cannot be determined simply from this criterion, because these experiments have significantly different initial conditions from those of the giant impact. An impact simulation needs to be performed with a new EOS that includes these experimental results.  

In addition, the real mantle material is not pure forsterite. Rather, perovskite is the dominant phase in the lower mantle ($P>$24 GPa).
Since perovskite is less compressible than forsterite (e.g., \citealt{JacksonAhrens1979, Dengetal2008}, Sarah Stewart, personal communications), material initially in the perovskite phase may be less shock-heated by an impact than our pure forsterite mantle.
However, we conclude that this has a minor effect on the outcome, at least for the thermodynamics of the particles that end up in the disk.
First, the larger part of the disk particles was originally in the upper mantle. 
The fractions of the disk particles that are from the upper-mantle are (a) standard 79\%, (b) fast-spinning Earth 60\%, (c) sub-Earths 68\%, (d) intermediate 42\%.
Additionally, most of the disk particles suffer additional shocks in which the pre-shock pressure is much lower than 24 GPa.
These multiple shocks efficiently increase the entropy of the disk particles and erase the memory of the initial condition.
Therefore, the choice of the phase in the mantle is not likely to affect the entropy estimate significantly.
We are separately considering the difficult question of the outcome of these events for the deep Earth.

%% file: conclusions.tex
This is the first work that bridges a hydrodynamic giant impact simulation and the resulting hydrostatic disk. We perform various SPH simulations in order to identify the properties of the disk, including its mass, angular momentum and entropy distribution. 
Using these values as constraints, the two dimensional structure of the disk is derived. 
Four distinctive scenarios are investigated: (a) standard, (b) fast-spinning Earth, (c) sub-Earths, and (d) intermediate. 
In all cases, the disk is approximately isentropic. In (a) and (d), the temperature of the disk is up to 4500-5000K and the overall vapor mass fraction of the disk is 20-30\%. These results are consistent with previous studies on the disk. On the other hand, the recently suggested models, such as (b) and (c), create a high-entropy and more vaporized disk. The temperature is as high as 6000-7000K and the vapor mass fraction is higher than 80\%. 
Such a high-entropy disk might lead to a chemically and isotopically different Moon from that of the canonical model.
We also develop semi-analytic solutions for the thermal structure of the disk, including the radial temperature distribution on the mid-plane and the radial vapor mass fraction. 
This model only requires the radial surface density and the average entropy of the disk as inputs. 
This may be used as an initial condition for further study on the Moon-forming disk.

%% file: appendix.tex
Two additional processes that heat up the disk, besides the impact-induced shock heating, are considered here.
As described in Section \ref{model:disk}, we assume that an initially eccentric and inclined disk particle is circularized.
Since this lowers the orbital energy of the particle, the excess of  the energy is emitted as heat and heats up the disk.
Assuming the component of angular momentum which is parallel to the Earth's spin axis is conserved, the resulting semi-major axis is $a_{i,  \rm final}=a_i (1-e_i^2) \cos^2 I_i$. $a_i$ is the initial semi-major axis before the damping, $e_i$ is the eccentricity, and $I_i$ is the inclination. 
Additional heating $\Delta E_i$ due to the circularization can be expressed as  
\begin{equation}
\Delta E_i=\frac{GM_{\oplus}m_i}{2a_i}\left(1-\frac{1}{(1-e_i^2) \cos^2 I_i}\right).
\label{eq:energy}
\end{equation} 
The entropy increase by this process $\Delta S_{i, \rm circular}$ is $\sim \Delta E_i/T_i$, where $T_i$ is the temperature of the particle.
In addition to this, in the vapor-rich cases, the disk is heated due to a mass redistribution within the disk, as described in Section \ref{model:disk}. Similarly, let $\Delta U$ equal the difference of the potential energies of the disks before and after the redistribution. The additional entropy is approximately written as $\Delta S \sim \Delta U/T_{\rm ave}$, where $T_{\rm ave}$ is the average temperature of the disk.
The increments of $S_{\rm ave}$ by these post-impact processes depend on the initial conditions of the impact.
In our calculations, each process increases the entropy by less than 10 percent.

%% file: acknowledgement.tex
We would like to thank Hidenori Genda for the helpful discussions about inherent problems of SPH, Kaveh Pahlevan, Aaron Wolf, Sarah Stewart, Robin Canup and anonymous reviewers for the insightful comments, Jay Melosh for sharing ANEOS data, and Takaaki Takeda for providing a visualization software, Zindaiji 3.
Numerical computations were carried out on GRAPE system at Center for Computational Astrophysics, National Astronomical Observatory of Japan.